\documentclass[12pt,preprint]{aastex}
\begin{document}
\title{The $3.6-8.0$ \micron~Broadband Emission Spectrum of HD 209458b: Evidence for an Atmospheric Temperature Inversion}
\author{Heather A. Knutson, David Charbonneau\altaffilmark{1}, Lori E. Allen}
\affil{Harvard-Smithsonian Center for Astrophysics, 60 Garden Street, Cambridge, MA 02138}
\altaffiltext{1}{Alfred P. Sloan Research Fellow}
\email{hknutson@cfa.harvard.edu, dcharbonneau@cfa.harvard.edu, leallen@cfa.harvard.edu, }

\author{Adam Burrows}
\affil{Steward Observatory, The University of Arizona, 933 North Cherry Avenue, Tucson, AZ 85721}
\email{burrows@zenith.as.arizona.edu}

\and

\author{S. Thomas Megeath}
\affil{Department of Physics and Astronomy, University of Toledo, 2801 West Bancroft Street, Toledo, OH 43606}
\email{megeath@astro1.panet.utoledo.edu}

\begin{abstract}

We estimate the strength of the bandpass-integrated thermal emission from the extrasolar planet HD 209458b at 3.6, 4.5, 5.8, and 8.0~\micron~using the Infrared Array Camera (IRAC) on the \emph{Spitzer Space Telescope}.  We observe a single secondary eclipse simultaneously in all four bandpasses and find relative eclipse depths of $0.00094\pm0.00009$, $0.00213\pm0.00015$, $0.00301\pm0.00043$, and $0.00240\pm0.00026$, respectively.  These eclipse depths reveal that the shape of the inferred emission spectrum for the planet differs significantly from the predictions of standard atmosphere models; instead the most plausible explanation would require the presence of an inversion layer high in the atmosphere leading to significant water emission in the 4.5 and 5.8~\micron~bandpasses.  This is the first clear indication of such a temperature inversion in the atmosphere of a hot Jupiter, as previous observations of other planets appeared to be in reasonably good agreement with the predictions of models without such an inversion layer.
                                                        
\end{abstract}

\keywords{infrared: techniques: photometric - eclipses - stars:individual: HD 209458b - planetary systems}

\section{Introduction}\label{intro}

''Hot Jupiters'' are a class of gas giant planets orbiting extremely close to their parent stars.  Twenty one of these planets have been observed to transit their parent stars \citep{mcc06,odon06,odon07,bak07a,bak07b,coll07,burk07,char07a,mand07,torr07}, allowing us to estimate not only their masses but also radii, temperatures, and other properties.  Of these twenty one systems, fifteen are bright enough for the type of \emph{Spitzer} observations described in this paper, including: HAT-P-1,2, and 3, HD 189733, HD 209458, HD 149026, TrES-1, 2, 3, and 4, WASP-1 and 2, and XO-1, 2, and 3. Of the fainter systems, including the five OGLE planets and CoRoT-Exo-1, most if not all (the position of CoRoT-Exo-1 has not yet been announced) lie in crowded fields where it is possible to achieve high quality infrared relative photometry using large-aperture ground-based telescopes and nearby comparison stars, as demonstrated by \citet{snell07}.  The radii of most of these planets are well-described by basic models of an irradiated hot Jupiter \citep{bod01,show02,bod03,bar03,laugh05a,bur07a}, but there are currently four planets (HD 209458b, WASP-1b, TrES-2, and TrES-4) with radii that appear to be significantly larger than predicted \citep{odon06,knut07a,char07b,shpo07,mand07}.  TrES-4 is the most extreme example, with an average density of only $0.222\pm0.045$~g cm$^{-3}$ \citep{mand07}.  The planet HAT-P-1b \citep{bak07a} was initially thought to share this property as well, but subsequent observations \citep{winn07} revealed that its radius was smaller than initially estimated.

Given that there is a clear distinction between hot Jupiters with inflated radii and those with normal (i.e. consistent with models) radii, a comparative study of the infrared emission spectra of the planets in these two classes might reveal important differences.  There are currently published estimates of the infrared emission from five planets, although this sample will expand significantly in the near future.  Of the planets with normal radii, TrES-1 has been observed at 4.5 and 8.0 \micron~\citep{char05}, and HD 189733b has been observed at 8.0~\micron, $7.5-14.7$~\micron, and 16~\micron~\citep{dem06,grill07,knut07b}.  HD 209458b, the only planet with an inflated radius in this sample, has been observed at $7.5-13.2$~\micron~and 24~\micron~\citep{dem05,rich07}.  Lastly, the core-dominated hot Jupiter HD 149026b \citep{har07} and the hot Neptune GJ 436b \citep{dem07b,demory07} have both been observed at 8.0 \micron.  This last planet is of particular interest, both because of its small size and significantly lower 8.0~\micron~brightness temperature of $\sim 710$~K.  It is likely that this planet has an atmospheric composition significantly different than its hotter and more massive cousins, which should be revealed by future \emph{Spitzer} observations at additional wavelengths.

Based on these initial results, it appears that the dayside emission spectra for the non-inflated hot Jupiters (excluding the core-dominated HD 149026b) are broadly consistent with the predictions of standard cloud-free atmosphere models \citep{sud03,seag05,bar05,fort05,bur06}.  The single exception is the $7.5-14.7$~\micron~IRS spectrum of HD~189733b measured by \citet{grill07}, which did not appear to have the predicted water absorption feature at the shortest wavelengths.  However, \citet{fort06} suggested that this planet might have a close-to-isothermal temperature profile on the dayside, which would wash out any absorption bands.  \citet{fort07} note that the broadband 8~\micron~eclipse depth for this planet appears to be deeper than the comparable eclipse depth from the IRS spectrum at these wavelengths, which suggests imperfect instrumental calibration of the IRS specra.  Intriguingly, the $7.5-13.2$~\micron~IRS spectrum of the inflated planet HD~209458b also shows no evidence of a water absorption feature at the shortest wavelengths; in addition, this spectrum appears to contain two \emph{emission} features, one of which has been tentatively identified as emission from silicate clouds \citep{rich07}.  The presence of high-altitude clouds in the atmosphere of HD~209458b would also explain the weaker-than-predicted sodium absorption in the planet's transmission spectrum \citep{char02}, but there is currently no definitive evidence for the presence of such clouds.

In this paper we present the first estimates of the infrared emission spectrum of a planet from the ``inflated'' category of hot Jupiters at wavelengths shorter than 7.5~\micron.  Atmosphere models for these planets  predict that many of the strongest features from CO, H$_2$O, and CH$_4$ will be located at wavelengths shorter than 8.0~\micron, making this a particularly interesting wavelength region to study.  The higher precision achievable in the $3-8$~\micron~wavelength range and complementary multi-wavelength information allows us to test the predictions of atmosphere models for this planet, including the pressure-temperature profile of the dayside atmosphere and the relative strength of potential emission and/or absorption features in the planet's spectrum from CO, H$_2$O, and CH$_4$. 

\section{Observations}\label{obs}

We observed HD 209458b over a period of 8.1  hours on UT 2005 Nov. 27, spanning a single secondary eclipse, using the Infrared Array Camera (IRAC) \citep{faz04} on the \emph{Spitzer Space Telescope} \citep{wern04}.  We observed in subarray mode with an exposure time of 0.1~s and cycled between the four IRAC channels in order to obtain estimates of the depth of the eclipse at 3.6, 4.5, 5.8, and 8.0 \micron~simultaneously.  We obtained a total of 35,840 $32 \times 32$ pixel images in each channel\footnote{We use images processed using version S13.0 of the standard Spitzer pipeline, to avoid the additional noise introduced by a darkdrift correction that was applied to all subarray images beginning with version S14.0 of the pipeline, released in May 2006.  This darkdrift correction is poorly constrained for subarray images dominated by a single bright star, and as a result introduces noise at a level higher than the effect it is meant to correct.  See IRAC pipeline history available online at http://ssc.spitzer.caltech.edu/irac/ for more information.}.  Images are taken in sets of 64, and four sets of images ($4 \times 64$ images total) are obtained in each channel before repointing the telescope to position the star correctly on the subarray for the next channel.  The position of the star on the subarray is still varying through the first set of 64 images after each repointing, and we chose to discard this initial set of images, leaving a total of 26,880 usable images in each of the four channels.  The total size of this pointing drift is 0.3 pixels in the first set of 64 images, and 0.1 pixels or less in the following three sets of images.

Because the two shortest wavelength IRAC channels (3.6 and 4.5~\micron) use InSb detectors and the two longer wavelength channels (5.8 and 8.0~\micron) use Si:As detectors, there are fundamental differences between the properties of the data taken with these two types of detectors.  We describe our analysis for each type of detector separately below.

\subsection{3.6 and 4.5~\micron~Observations (InSb Detector)}\label{short_norm}

Because HD 209458 is a bright star ($K=6.31$) and the background at these shorter wavelengths is minimal, we calculate the flux from the star in each image using aperture photometry with a radius of five pixels.  We determine the position of the star in each image as the position-weighted sum of the flux in a $7\times7$ pixel box centered on the approximate position of the star.  We estimate the background in each image by selecting a subset of pixels from the corners of the image where the point spread function of the star is faintest, making a histogram of the flux values in these pixels, and fitting a Gaussian function to the center of this distribution.  We calculate the JD value for each image as the time at mid-exposure, and apply a correction to convert these JD values to the appropriate HJD, taking into account Spitzer's orbital position at each point during the observations.  As a check we repeat our analysis using apertures ranging from $3.5-7$ pixels, and obtain consistent results in all cases.

Fluxes measured at these two wavelengths show a strong correlation with the changing position of the star on the array, at a level comparable to the depth of the secondary eclipse.  This effect is due to a well-documented intra-pixel sensitivity \citep{reach05,char05,mor06}, and can be removed by fitting the data with a quadratic function of x and y position, where position is measured as the distance between the peak of the star's point spread function and the center of the pixel containing this peak:

\begin{equation}
f^1=f*(c_1+c_2*(x-14.5)+c_3*(x-14.5)^2+c_4*(y-14.5)+c_5*(y-14.5)^2)
\end{equation}

where $f$ is the original flux from the star, $f^1$ is the measured flux, $x$ and $y$ denote the location of the center of the star on the array, and $c_1-c_5$ are the five free parameters in the fit.  We find that adding higher-order terms to this equation does not improve the fit, nor does adding a linear or quadratic function of time.  We fit this function to the out-of-transit data alone and also simultaneously with the transit curve, and obtain consistent results in both cases.  We chose to use the simultaneous fit, as it allows us to accurately estimate the additional uncertainty in the depth of the eclipse introduced by this correction.

\begin{figure}
\epsscale{0.7}
\plotone{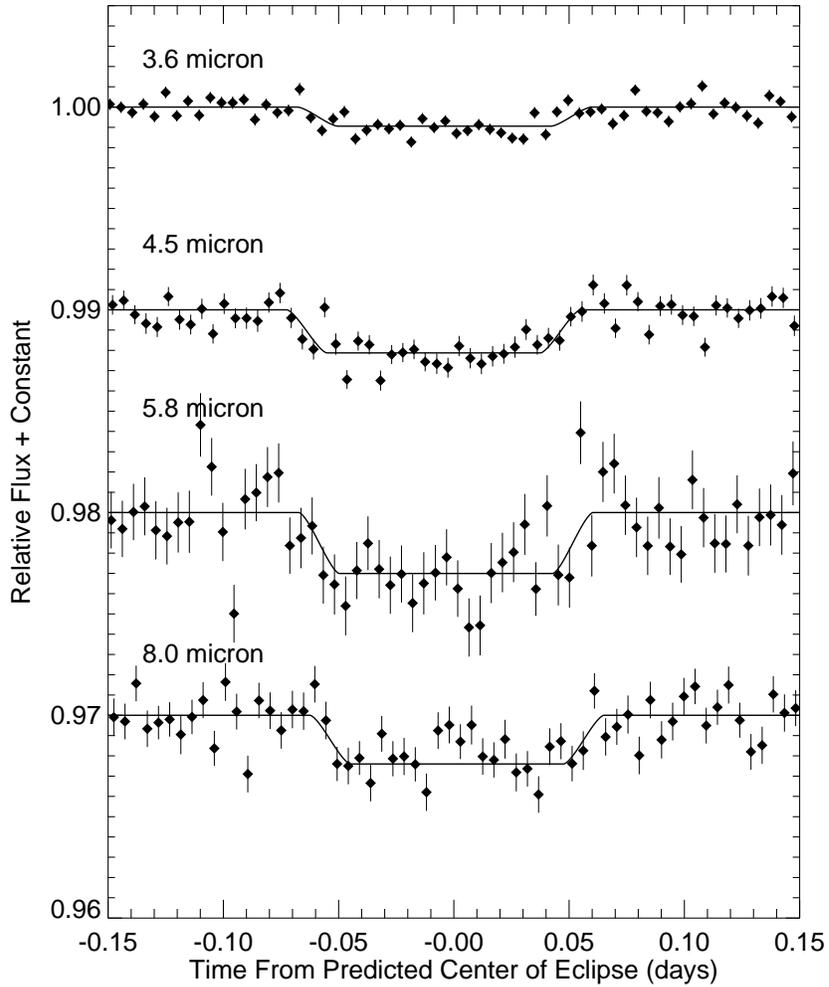}
\caption{Secondary eclipse of HD 209458b on UT 2005 Nov. 27, observed at (from top to bottom) 3.6, 4.5, 5.8, and 8.0 \micron, with best-fit eclipse curves overplotted.  Data has been normalized to remove detector effects (see discussion in \S\ref{short_norm} and \S\ref{long_norm}), and binned in 7 minute intervals.\label{four_eclipses}}
\end{figure}

We fit the correction for the intra-pixel sensitivity of the array and the transit curve simultaneously to the data using a Markov Chain Monte Carlo method \citep{ford05,winn07} with $10^6$ steps.  We set the uncertainty on individual points equal to the standard deviation of the out-of-transit data after correction for the intra-pixel variations, and remove outliers of 5$\sigma$~or more as calculated using the residuals from the best-fit light curve.  We allow both the depth and timing of the secondary eclipse to vary independently for the eclipses at each of the two observed wavelengths, and take the other parameters for the system (planetary and stellar radii, orbital period, etc.) from \citet{knut07a}.  We calculate our transit curve using the equations from \citet{mand02} for the case with no limb-darkening.  After running the chain, we search for the point in the chain where the $\chi^2$ value first falls below the median of all the $\chi^2$ values in the chain (i.e. where the code had first found the best-fit solution), and discard all the steps up to that point.  We take the median of the remaining distribution as our best-fit parameter, with errors calculated as the symmetric range about the median containing 68\% of the points in the distribution.  The distribution of values was very close to symmetric in all cases, and there did not appear to be any strong correlations between variables.  Figure \ref{norm_plots} shows the binned data with the best fit to the detector effects overplotted, and Figure \ref{four_eclipses} shows the binned data once these trends are removed, with best-fit eclipse curves overplotted.  Best-fit eclipse depths and times are given in Table \ref{eclipse_depths}.

\subsection{5.8 and 8.0~\micron~Observations (Si:As Detector)}\label{long_norm}

At longer wavelengths the flux from the star is smaller and the zodiacal background is larger; as a result we chose to use a smaller 3.5 pixel aperture at these two wavelengths in order to minimize the noise contribution from this increased background.  As a test we also tried using a psf fit to derive the time series in the 8~\micron~channel, which has the highest background, using the in-flight point response functions generated from calibration test data \footnote{Available at http://ssc.spitzer.caltech.edu/irac/psf.html}.  There was no improvement in the resulting time series, indicating that aperture photometry is still appropriate here.  As a check we repeat our analysis using apertures ranging from $3.5-5$ pixels and obtain a consistent signal in all cases, but with a scatter that increases with the radius of the photometric aperture.  As before, we calculate the position of the star individually in each image as the position-weighted sum of the fluxes in a $7\times7$ pixel box, and estimate the background  using a Gaussian fit to a histogram of the pixels in the corners of the array.  Fluxes in the first 10 images in each set of 64 are consistently below the median value for the set by as much as 10\%, with the lowest values at the beginning of each set, so we chose to exclude the first 10 images from each set of 64 in our analysis.  

There is no intra-pixel sensitivity at these wavelengths, but there is another well-documented detector effect \citep{knut07b} which causes the effective gain (and thus the measured flux) in individual pixels to increase over time.  This effect has been referred to as the ``detector ramp'', and has also been observed to occur in the IRS and MIPS 24~\micron~arrays, which are made from the same material \citep{dem05,dem06}.  The size of this effect depends on the illumination level of the individual pixel; pixels with high ($>$250 MJy Sr$^-1$ in the 8~\micron~channel) will converge to a constant value within the first hour of observations, while lower-illumination pixels will show a linear increase in the measured flux over time, with a slope that varies inversely with the logarithm of the illumination level.

This effect is important for two reasons.  First, it means that the observed 3\% linear increase in the measured background flux at 8~\micron~over the period of the observations is most likely not the result of a real change in the zodiacal background, but is instead another example of this detector ramp. Although the increased noise and smaller size of the background at 5.8~\micron~obscures this effect, there appears to be a similar upward trend.  Thus, rather than calculating the background in each image individually and subtracting that value, we subtract a constant background of 3.21 MJy Sr$^-1$ per pixel from all the 8~\micron~images, and 0.41 MJy Sr$^{-1}$ per pixel from all 5.8~\micron~images.  This background is calculated as the median background value during the last 2.5 hours of observation, when presumably the background is closest to its true value.  We note that this choice has a negligible effect on our final eclipse depths, as the background constitutes only 0.2\% and 2.3\% of the signal in our aperture at 5.8 and 8.0~\micron, respectively.  If we subtract no background at all, the eclipse depths we obtain are still well within the 1$\sigma$~uncertainties of our final quoted values.

\begin{figure}
\epsscale{0.7}
\plotone{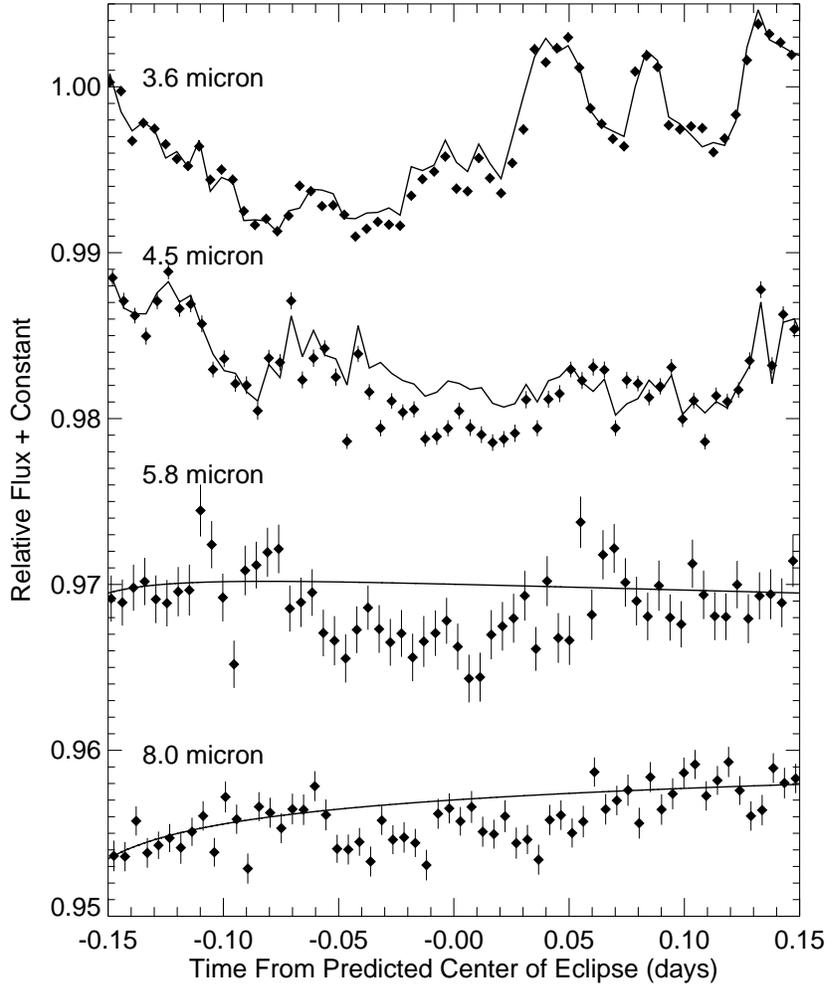}
\caption{Secondary eclipse of HD 209458b on UT 2005 Nov. 27, observed at (from top to bottom) 3.6, 4.5, 5.8, and 8.0 \micron, binned in 7 minute intervals and normalized to one.  The overplotted curves show the best-fit corrections for detector effects (see \S\ref{short_norm} and \S\ref{long_norm}). \label{norm_plots}}
\end{figure}

This effect also produces a 0.5\% increase in the measured flux from the star at 8.0~\micron~over the period of these observations (see Figure \ref{norm_plots}), and a much smaller ($<$ 0.1\%) increase at 5.6~\micron.  Unlike the detector ramp at low illumination levels, the ramp for higher illuminations has an asymptotic shape, with a steeper rise in the first 30 minutes of observations.  We discard the first 30 minutes of data in both the 5.8 and 8.0~\micron~channels and fit the remaining binned time series from our 3.5 pixel aperture with a quadratic function of ln(dt), where dt is the change in time from the start of the observations.  \citet{knut07b} used this same functional form (with additional degrees of freedom) to describe the detector ramp in the 8~\micron~channel over a period of 33 hours of continuous observations of HD~189733, and it accurately captures the behavior of this ramp for a range of illumination levels.  Unlike \citet{knut07b} we do not attempt to correct each of the pixels in the images individually for this ramp; this is not neccessary for our analysis, and the lower fluxes, shorter time frame (8 hours instead of 33) and reduced cadence of our data (from cycling between the four detectors) make it difficult to characterize this effect accurately at the pixel level.  Instead we assume the detector ramp in the binned flux in our 3.5 pixel aperture will have a shape similar to the ramp for individual pixels and fit the binned time series in each channel with a quadratic function of ln(dt), which produces a good fit to the observed trends.

We fit both the quadratic function of ln(dt)  and the transit curve to the data simultaneously using a Markov Chain Monte Carlo method as described in \S\ref{short_norm}.  As before, the distribution of values was very close to symmetric in all cases, and there did not appear to be any strong correlations between variables.  Best-fit eclipse depths and times from these fits are given in Table \ref{eclipse_depths}, and the time series both before and after correcting for detector effects are shown in Figures  \ref{norm_plots} and \ref{four_eclipses}, respectively.  As a check we repeated these fits using both a linear function in time and a linear function of ln(dt), and found that the eclipse depth in both cases varied by less than 1$\sigma$ from our quoted value.

\section{Discussion}

We ultimately achieve noise levels in each of the four bandpasses that are 1.5, 1.9, 2.0, and 1.6 times higher than the predicted photon noise at 3.6, 4.5, 5.8, and 8.0~\micron, respectively.  The additional noise is most likely the result of the jitter introduced by the need to repoint the telescope every 3.5 minutes in order to switch between bandpasses.  Although the eclipse is detected to a high degree of significance in all four bandpasses, it is worth noting that we ultimately achieve a precision at 4.5 and 8.0~\micron~comparable to that of the measured secondary eclipse depths for TrES-1 \citep{char05}, even though this star is significantly fainter ($K=9.82$ vs $K=6.32$ for HD 209458).  This is partially explained by the reduced cadence of the HD 209458 observations, which had 15\% of the total effective integration time per band for the TrES-1 observations, but the frequent repointings appear to have contributed additional noise as well.  We also note that \citet{char05} did not include the uncertainties contributed by their fits to the trends in the out-of-transit data in their error estimates.

If we combine our estimates of the eclipse timing in each of the four bandpasses, we find that the center of the eclipse occurs at $2453702.5251\pm0.0012$~HJD.  This is $4.9\pm1.7$~minutes or 2.9$\sigma$ earlier than predicted \citep{knut07a}, and this is without accounting for the additional 50~s delay in the predicted time due to the light travel time in the system \citep{loeb05}.  In this case the uncertainty in the predicted time is negligible compared to the uncertainty in our measurement.

It is possible that we have under-estimated the uncertainties in our estimates for the timing of the eclipse.  Because the ingress and egress occur on relatively short time scales, fitted values for the eclipse times are particularly sensitive to the presence of correlated noise in the time series, which is not included in the error bars from the Markov fits described above.  To estimate the effect of correlated noise on our best-fit eclipse depths and times, we use the ``prayer-bead'' method as described in \citet{gill07}. In this method we take the time series of the residuals from our best-fit solution for each eclipse, shift the residuals forward in time with the points at the end of the time series wrapping back around to the beginning, add the best-fit solution back in, and fit the new timeseries for the full set of parameters.  This process is iterated until the residuals have been shifted back around to their original positions in the timeseries.  The variance in the resulting set of values for the best-fit eclipse depths and transit times gives the error values for each of the parameters.  In this case we set our uncertainties equal to the range in values containing 68\% of the points in the distribution in a symmetric range about the median for a given parameter, as we did before with our Markov fits in \S\ref{short_norm}.  The resulting uncertainties are comparable to the uncertainties we obtained from our Markov fits, with some larger and some smaller values.  For the eclipse depths, we find uncertainties of [0.0009,~0.00025,~0.00072,~0.00020] at [3.6,~4.5,~5.8,~8.0]~\micron.  For the best-fit eclipse times, we find uncertainties of [4.0,~2.9,~5.4,~1.9] minutes.  From this we conclude that correlated noise is not a significant source of uncertainty in the data.  Significantly, we note that the uncertainty in the best-fit time for the 4.5~\micron~eclipse, which occurs 12.6 minutes earlier than predicted, is \emph{smaller} than the original uncertainty from our Markov fit. We elect to use the uncertainties from our Markov fits as the final uncertainties for our parameters, as we feel that this is the better method in the case where correlated noise in the data is minimal, as it appears to be.  In either case the differences between the two methods for calculating uncertainties are minor and do not affect our conclusions.

Although the combined best-fit time for the four eclipses appears to occur 4.9 minutes or 2.9$\sigma$ earlier than predicted, we do not believe that this is a convincing detection of a non-zero orbital eccentricity.  Because we observe the eclipse simultaneously at four wavelengths, we would expect to see similar timing offsets in all four channels if the shift was the result of a non-zero eccentricity.  Instead, there appears to be a larger, marginally significant (3.6$\sigma$) offset in the 4.5~\micron~channel, while the other three channels are effectively consistent with zero offset (see Table \ref{eclipse_depths}).  A more plausible explanation would be an apparent timing shift caused by a color-dependent non-uniform brightness distribution on the surface of the planet, which would alter the shape of the ingress and egress relative to the shape expected for a uniform brightness distribution \citep{will06,rau07}.  It is reasonable to expect that this brightness distribution might vary with wavelength, causing different apparent timing shifts in each of the four channels, as different wavelengths probe different depths in the atmosphere.  The signal-to-noise of our data is not high enough to distinguish the changes in the shape of the ingress and egress that would signal such a non-uniform brightness distribution, but continuous, higher-cadence observations similar to the 8~\micron~observations of HD 189733b by \citet{knut07b} might provide a definitive answer to this question in the future.

\begin{figure}
\epsscale{0.9}
\plotone{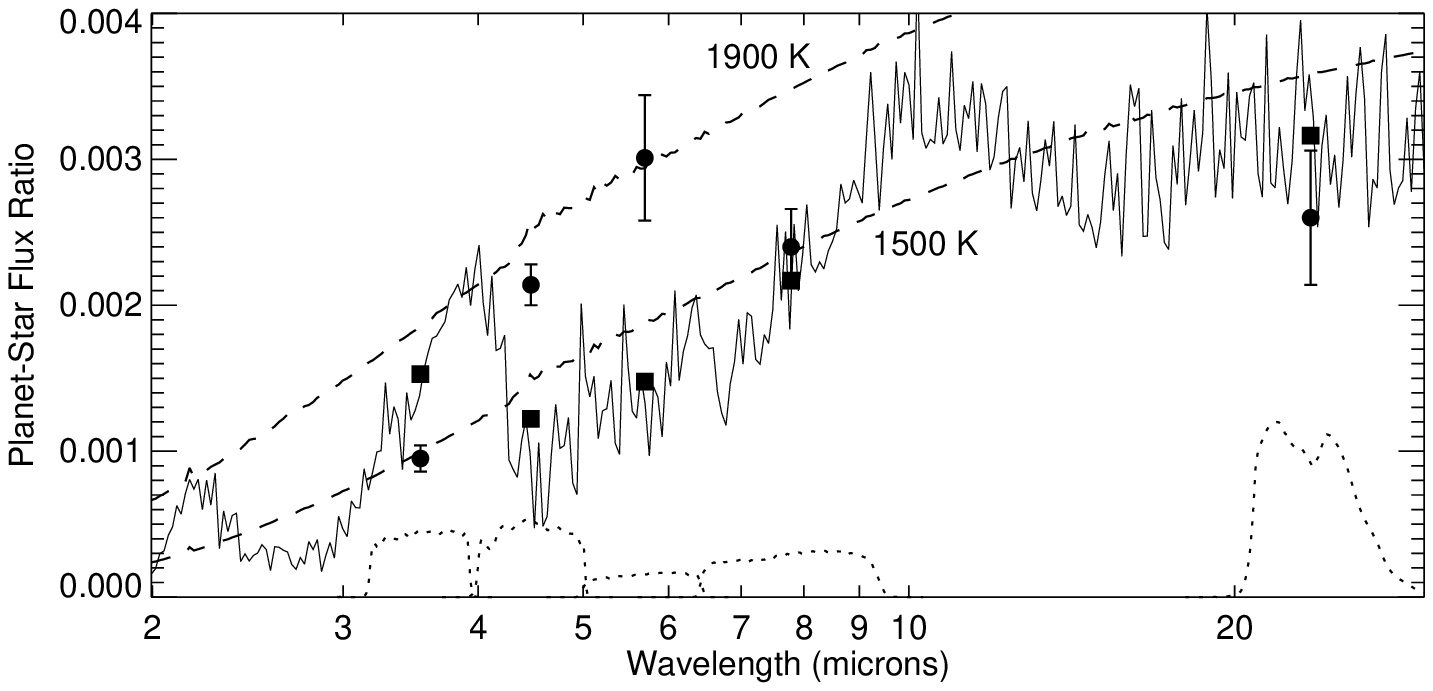}
\caption{Predicted emission spectrum for HD 209458b from \citet{bur06}.  Squares show this spectrum integrated over the \emph{Spitzer} bandpasses (response functions for these bandpasses are shown at the bottom of plot (dotted lines), scaled by a factor of 0.001), and circles are the estimated eclipse depths in these bandpasses.  The 24~\micron~point is taken from \citet{dem05}.  The two dashed lines give the planet-star flux ratio for the case where the planet is a perfect blackbody with a temperature of either 1500~K or 1900~K and the spectrum of the star is calculated from a model (available at http://kurucz.harvard.edu/stars/hd209458).  Note that we have chosen not to plot the $7.5-12.2$~\micron~spectrum~measured by \citet{rich07}; this is because the spectrum was scaled to match a preliminary value for our 8.0~\micron~eclipse depth, and thus does not contain independent information about the absolute strength of the emission from the planet at these wavelengths.\label{spectrum}}
\end{figure}

When we compare the measured eclipse depths at 3.6, 4.5, 5.8, and 8.0~\micron~to the predictions of a theoretical model for this planet (see Figure \ref{spectrum}), it is immediately clear that the shape of the observed spectrum differs significantly from the predicted values. The data in Fig. \ref{spectrum} show a peak in flux centered around the $5.8$~\micron~bandpass and indicate that the flux in the 4.5~\micron~bandpass exceeds that in the 3.6~\micron~bandpass.  As Fig. \ref{spectrum} suggests, the effective photospheres of the 4.5 and 5.8~\micron~features are at rather high temperatures in the atmosphere and the corresponding effective temperature of the 3.6~\micron~flux is at a lower temperature. Previous theory \citep{bur05,fort05,bar05,seag05} had suggested that there would be a trough between the 3.6 and 8.0~\micron~IRAC bandpasses due to a water absorption feature and that the flux at 3.6~\micron~would either exceed that at 4.5~\micron~or would be comparable to it.  However, the new IRAC data force us to conclude that there is a thermal inversion in the atmosphere of HD~209458b at altitude and that we are indeed seeing water in the $4-8$~\micron~gap, but in {\it emission}. In fact, we now expect the water absorption features of the older default theory to be flipped into emission features throughout the entire near- to mid-infrared wavelength range.  Therefore, we find that, contrary to the conclusion of \citet{rich07}, the flatness or slight rise of their IRS spectrum near $\sim$7.8~\micron~in fact supports the presence of abundant atmospheric water, because this spectral region is at the edge of a strong water band in emission.  A temperature inversion might also naturally explain the two emission features tentatively identified by \citet{rich07}.  \citet{bur07b} explore the theoretical and model consequences of these new IRAC data in more depth. 

The idea that the spectrum of a strongly irradiated extrasolar giant planet could manifest water emission features was presaged by \citet{hub03} and \citet{bur06}, who discussed a bifurcation in the atmosphere solution, but we still do not know the nature of the stratospheric absorber responsible for this heating.  \citet{rowe06,rowe07} report a value of $0.04\pm0.04$ for the geometric albedo of the planet at visible wavelengths, indicating that the stratospheric absorber must absorb much more in the optical than the infrared, as predicted by \citet{hub03}. This eliminates many types of clouds, which tend to be reflective at optical wavelengths.  \citet{cow07} placed a 2$\sigma$~upper limit of 0.0015 on the size of the phase variation for this planet at 8~\micron; when compared to the eclipse depth of $0.00240\pm0.00026$ described in this paper, this indicates that the flux from the night side must be at least 60\% of the flux from the day side, which would indicate relatively efficient circulation in the atmosphere at the level of the 8~\micron~photosphere.  Because high-altitude clouds or other opaque layers would shift the location of the 8~\micron~photosphere, this provides additional constraints on the nature of the unknown stratospheric absorber. 

\section{Conclusions}

We estimate the secondary eclipse depth for the transiting planet HD~209458b at 3.6, 4.5, 5.8, and 8.0~\micron.  These observations provide a useful complement to previous observations of this system, which were limited to wavelengths longer than 8.0~\micron.  In contrast to the results for longer wavelengths, we find that the planet's emission at shorter wavelengths is clearly inconsistent with the predictions for a standard cloudless atmosphere model.  We suggest an alternative explanation, in which a temperature inversion in the upper atmosphere produces water emission features at 4.5 and 5.8~\micron.  Although the cause of this inversion layer is unknown, it is suggestive that this planet also falls into a class of hot Jupiters which appear to have radii that are significantly larger than predicted by standard models for an irradiated gas giant.  These two characteristics may or may not be related; the obvious next test would be to extend these observations to a much larger sample of planets to determine which, if any, of the other 21 known transiting planets show similar temperature inversions.  The very bright 8.0~\micron~flux from the core-dominated planet HD~149026b \citep{har07} indicates that this planet may also have a temperature inversion, as predicted by \citet{fort06b}; this result should be confirmed by \emph{Spitzer} observations at additional wavelengths in the near future.

Given the imminent depletion of \emph{Spitzer's} cryogen, at which point only the 3.6 and 4.5~\micron~channels will be functioning, it is worth noting that the best evidence for the temperature inversion in HD 209458b's atmosphere comes from observations in these two shorter-wavelength channels.  \emph{Spitzer} provides the optimal platform for this type of measurement, and it would be relatively straightforward to survey all of the known bright transiting systems as part of the post-cryogenic mission\footnote{See, for example, the white paper on the Warm Spitzer science prospects by John Stauffer and the white paper on exoplanets by D. Deming et al. from the Warm Spitzer Workshop, available at http://ssc.spitzer.caltech.edu/mtgs/warm/wp.html.}.  Such a survey has the potential to provide a definitive answer to the question of what properties of the planet or parent star lead to these temperature inversions, and perhaps shed light on the nature of the clouds or other upper-atmosphere optical absorbers that are needed to produce the temperature inversion.

\acknowledgements

This work is based on observations made with the \emph{Spitzer Space Telescope}, which is operated by the Jet Propulsion Laboratory, California Institude of Technology, under contract to NASA.  Support for this work was provided by NASA through an award issued by JPL/Caltech.  HAK was supported by a National Science Foundation Graduate Research Fellowship.  AB would like to acknowledge support from NASA under grants NNG04GL22G and NNX07AG80G and through the NASA Astrobiology Institute under Cooperative Agreement No. CAN-02-OSS-02 issued through the Office of Space Science.  We would also like to thank J. Matthews for sharing MOST results in advance of publication.

\clearpage

\begin{deluxetable}{lrrrrcrrrrr}
\tabletypesize{\scriptsize}
\tablecaption{Best-Fit Eclipse Depths and Times \label{eclipse_depths}}
\tablewidth{0pt}
\tablehead{
\colhead{$\lambda$ (\micron)} & \colhead{~~Relative Eclipse Depth}  & \colhead{Center of Transit (HJD)} & \colhead{O$-$C (min.)\tablenotemark{b}}}
\startdata
3.6 & $0.00094\pm0.00009$ & $2453702.5244\pm0.0024$ & $-5.9\pm3.4$\\
4.5 & $0.00213\pm0.00015$ & $2453702.5198\pm0.0024$ & $-12.6\pm3.5$\\
5.8 & $0.00301\pm0.00043$ & $2453702.5251\pm0.0026$ & $-4.9\pm3.8$\\
8.0 & $0.00240\pm0.00026$ & $2453702.5299\pm0.0022$ & $2.0\pm3.1$\\
24\tablenotemark{a} & $0.0026\pm0.00046$ & $2453346.5278\pm0.0049$ & $-1.6\pm7.1$\\
\enddata
\tablenotetext{a}{\citet{dem05}}
\tablenotetext{b}{Predicted transit time from \citet{knut07a}}
\end{deluxetable}

\clearpage

\end{document}